\documentclass[iop]{emulateapj}
\usepackage{epsfig}
\usepackage{apjfonts}
\usepackage{aas_macros}
\usepackage{amsmath}
\usepackage{graphicx}

\begin{document}

\title{A more stringent constraint on the mass ratio of binary neutron star merger GW170817}
\author{He Gao$^{1,*}$, Zhoujian Cao$^{1,*}$, Shunke Ai$^{1}$, and Bing Zhang$^{2,3,4}$}
\affiliation{
$^1$Department of Astronomy, Beijing Normal University, Beijing 100875, China; gaohe@bnu.edu.cn;zjcao@bnu.edu.cn
\\$^2$Department of Physics and Astronomy, University of Nevada Las Vegas, NV 89154, USA;\\
  $^3$Department of Astronomy, School of Physics, Peking University, Beijing 100871, China; \\
  $^4$Kavli Institute of Astronomy and Astrophysics, Peking University, Beijing 100871, China.}

\begin{abstract}
Recently, the LIGO-Virgo collaboration reported their first detection of gravitational wave (GW) signals from a low mass compact binary merger GW170817, which is most likely due to a double neutron star (NS) merger. With the GW signals only, the chirp mass of the binary is precisely constrained to $1.188^{+0.004}_{-0.002}~\rm{M_{\odot}}$, but the mass ratio is loosely constrained in the range $0.4-1$, so that a very rough estimation of the individual NS masses ($1.36~{\rm M_{\odot}}<M_1<2.26~\rm{M_{\odot}}$ and $0.86~{\rm M_{\odot}}<M_2<1.36~\rm{M_{\odot}}$) was obtained. Here we propose that if one can constrain the dynamical ejecta mass through performing kilonova modeling of the optical/IR data, by utilizing an empirical relation between the dynamical ejecta mass and the mass ratio of NS binaries, one may place a more stringent constraint on the mass ratio of the system. For instance, considering that the red ``kilonova'' component is powered by the dynamical ejecta, we reach a tight constraint on the mass ratio in the range of $0.46-0.59$. Alternatively, if the blue ``kilonova'' component is powered by the dynamical ejecta, the mass ratio would be constrained in the range of $0.53-0.67$. Overall, such a multi-messenger approach could narrow down the mass ratio of GW170817 system to the range of $0.46-0.67$, which gives a more precise estimation of the individual NS mass than pure GW signal analysis, i.e. $1.61~{\rm M_{\odot}}<M_1<2.11~{\rm M_{\odot}}$ and $0.90~{\rm M_{\odot}}<M_2<1.16~{\rm M_{\odot}}$. 
\end{abstract}


\section {Introduction}

The ground-breaking discoveries of gravitational waves (GWs) due to double black hole (BH-BH) mergers (GW150914, GW151226, GW170104 and GW170814, \citealt{abbott16a,abbott16b,abbott17,abbott17b}) by the LIGO-VIRGO Collaboration, especially the discovery of the double neutron star merger GW170817 as well as the multi-wavelength electromagnetic counterparts by many teams \citep[][and reference therein]{abbott17c}, opened a brand new era of GW-led multi-messenger astronomy. 

GW170817 was first detected online by the single-detector LIGO-Hanford, but was quickly confirmed by a re-analysis of the joint data from LIGO-Hanford, LIGO-Livingston and Virgo detectors with a high confidence level \citep{abbott17c}. After a coherent Bayesian analysis of the three-instrument data by invoking marginalisation over calibration uncertainties and waveform models of compact binary coalescence, the GW source is located in a region of 28 $\rm deg^2$ at a distance $40\pm8$ Mpc. The chirp mass of the binary system is estimated as $1.188^{+0.004}_{-0.002}~\rm{M_{\odot}}$. Under the prior of high dimensionless NS spin ($\chi<0.89$), the mass ratio of the two compact objects is bound to the range of $0.4-1.0$, so that the mass of these two objects are estimated as $1.36~\rm{M_{\odot}}<M_1<2.26~\rm{M_{\odot}}$ and $0.86~\rm{M_{\odot}}<M_2<1.36~\rm{M_{\odot}}$. Under the prior of low dimensionless NS spin ($\chi<0.05$), on the other hand, the mass ratio of two compact objects is bound to the range of $0.7-1.0$, so that the mass of these two objects are estimated as $1.36~\rm{M_{\odot}}<M_1<1.60~\rm{M_{\odot}}$ and $1.17~\rm{M_{\odot}}<M_2<1.36~\rm{M_{\odot}}$. These values of the component mass are consistent with a double neutron stars system \citep{abbott17d}. 

Along with GW170817, a weak short-duration gamma-ray burst (SGRB) 170817A was detected by Fermi-GBM \citep{goldstein17} and INTEGRAL \citep{savchenko17}, which coincides with the GW signal in both trigger time and direction. Around 11 hours after the GW170817 signal, an associated optical transient SSS17a was discovered in the galaxy NGC 4993 \citep{coulter17}. A large number of teams across the world contributed to the observation of this source using ground- and space-based telescopes \cite[][for details]{abbott17d}. These detections are not unexpected. In the literature, many associated EM counterparts have been proposed for NS-NS mergers, including a short gamma-ray burst and its afterglow emission \citep{eichler89,narayan92}, an optical/IR transient known as the kilonova \citep{li98,metzger10} and some other broad-band signals \citep[e.g.][]{nakar11,zhang13,gao13,yu13,metzgerpiro14}. The brightness of each EM counterpart depends on the details of the poorly known merger physics, especially the neutron star equation of state (EoS), and the outcome of the post-merger central remnant object \citep{lasky14,gao16}. 

Numerical simulations showed that an NS-NS merger could eject a fraction of the total mass in two different channels, forming a mildly isotropic, sub-relativistic ejecta \cite[][for a review]{metzger17}: {\em a dynamical ejecta} that is tidally ripped and dynamically launched during the merger \citep{rezzolla11,rosswog13,bauswein13,hotokezaka13}, and {\em a disk wind} launched from a hot accretion disk around the central black hole due to the neutrino emission or other mechanisms \citep{fernandez13,lei13,just15,wu16,siegel17,song17,ma17}. If an ejecta has a sufficient high neutron fraction to produce a robust abundance pattern for heavy nuclei with $A\gtrsim130$, including a modest fraction of elements with partially-filled f-shell valence shells such as those in the lanthanide and actinide groups, then its opacity can be an order of magnitude or more higher than the opacity of iron \citep{kasen13,tanaka13}. In this case, the kilonova emission would be ``red'' with emission peaking at the near IR wavelengths on a timescale of several days to a week \citep{kasen13,barnes13}. On the other hand, a less neutron-rich ejecta could only produce lighter r-process elements with $90<A<130$, being free of Lanthanide group elements, so it would produce a ``blue'' kilonova emission, peaking at the visual bands R and I, on a timescale of about 1 day at a level 2-3 magnitudes brighter than the Lanthanide-rich case \citep{metzger14}. Earlier models \citep[e.g.][]{metzger14} suggested that the dynamical ejecta likely powers a red kilonova while the disk wind likely powers a blue component near the polar region.

The optical/IR counterpart of GW170817 indeed showed two distinct blue and red components 
\citep{evans17,drout17,nicholl17,smartt17,arcavi17,kasen17,cowperthwaite17,kilpatrick17,villar17}. Fitting the light curve, one could obtain the properties for both components, such as their velocities, opacities, and especially the masses of the two components. The inferred mass of both components, especially that of the red component, seem to be larger than that predicted for dynamical ejecta in previous simulations. On the other hand, simulations of the disk wind ejection \citep{siegel17} showed significant mass ejection. As a result, one may not clearly identify the red component as originating from the dynamical ejecta. Indeed, the opposite view of interpreting the blue component as due to the dynamical ejecta exists in the literature \citep[e.g.][]{kasen17,nicholl17}.

Given a NS EoS, the mass of the dynamical ejecta essentially depends on the mass ratio of the NS binary, with higher mass ratio binaries ejecting greater quantities of the mass \citep{bauswein13,lehner16}. In this paper, we show that making use of the ejecta mass information, even with the uncertainty of their physical origin, could lead to an interesting constraint on the mass ratio of the NS binary, which could lead to a much better estimation of the neutron star mass.

\begin{figure*}[t]
\begin{center}
\begin{tabular}{ll}
\resizebox{80mm}{!}{\includegraphics[]{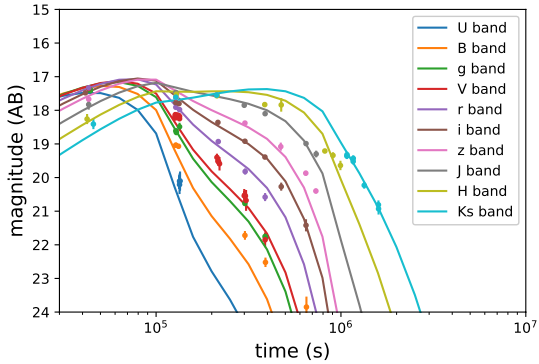}} &
\resizebox{80mm}{!}{\includegraphics[]{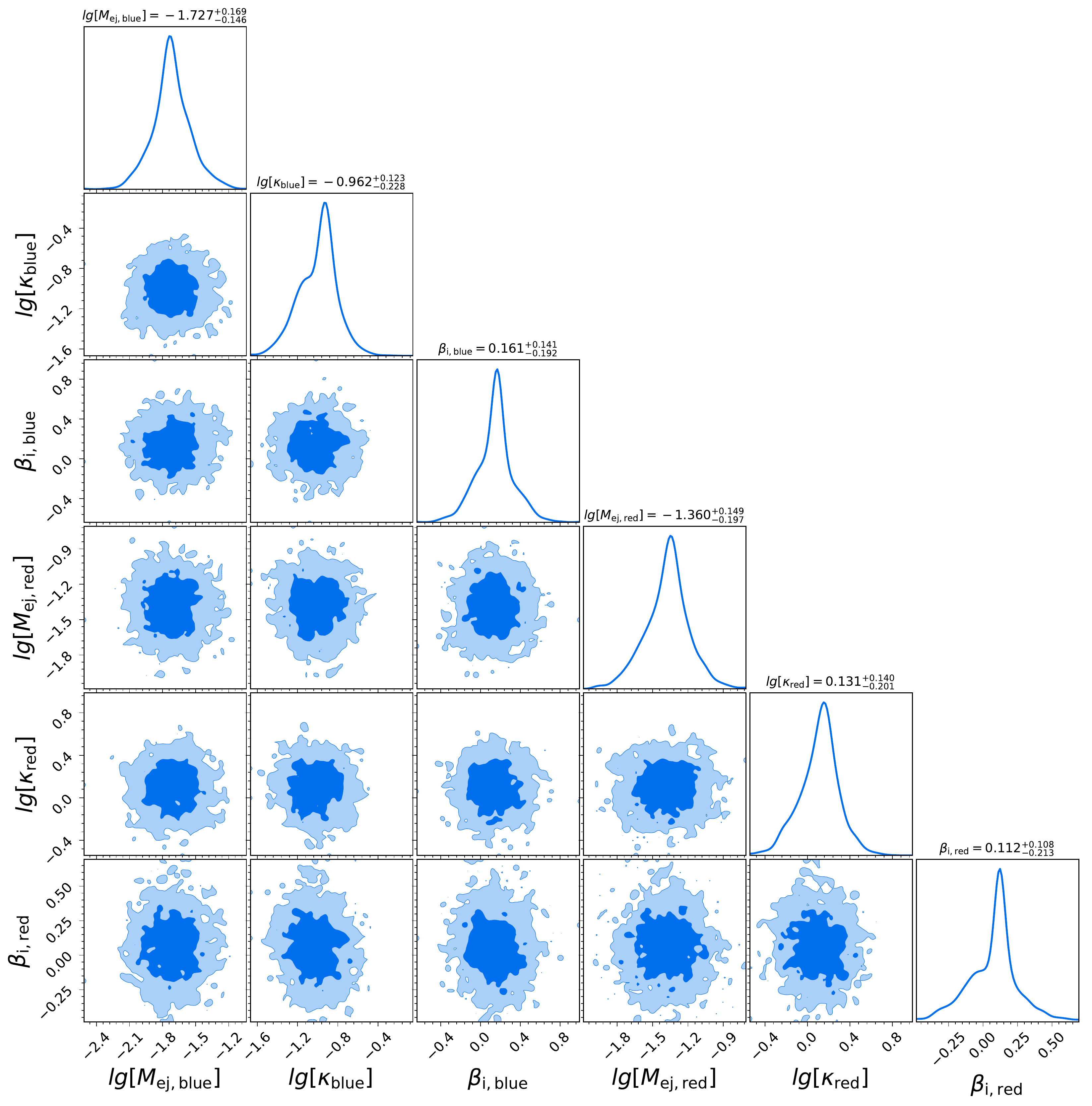}} \\
\end{tabular}
\caption{Left panel: multi-wavelength light curve data of the optical/IR counterpart of GW170817, along with the best fitting ``blue'' + ``red'' kilonova model. The data are taken from \cite{villar17}. Right panel: posterior probability plot of our model light curve fits.}
\label{fig:fit}
\end{center}
\end{figure*}

\section{Kilonova Modeling}

To interpret the data of the optical/IR counterpart of GW170817, we consider two components of neutron rich ejecta during an NS-NS merger: a ``red'' component with mass $M_{\rm ej,red}$, initial dimensionless speed $\beta_{\rm i, red}$, and electron fraction $Y_e\lesssim0.1$, contributing to a late ``red'' kilonova emission, and a ``blue'' component with mass $M_{\rm ej,blue}$, initial dimensionless speed $\beta_{\rm i, blue}$ and electron fraction $Y_e>0.3$, contributing to an early ``blue'' kilonova emission. Both ejecta receive heating from the radioactive decay of the heavy nuclei synthesized in the ejecta via the r-process. For both components, the dynamical evolution of the ejecta can be determined by\footnote{In principle, the ejecta may get accelerated into a (trans-)relativistic speed. Here we adopt a complete description of the dynamical evolution by covering both the non-relativistic and relativistic regimes. Since the peak luminosity of these transients can be much brighter than $10^{41} \ {\rm erg \ s^{-1}}$ which was used to define the term ``kilonova'' by \cite{metzger10}, such transients are also termed as ``merger-novae'' by \cite{yu13}.} \citep{yu13}
\begin{eqnarray}
{d\Gamma\over dt}={{\cal D}^{2}L'_{\rm ra}-{\cal D}^{2}L'_{\rm e}-\Gamma {\cal D}\left({dE'_{\rm int}\over
dt'}\right)-(\Gamma^2-1)c^2\left({dM_{\rm sw}\over dt}\right)\over
M_{\rm ej}c^2+E'_{\rm int}+2\Gamma M_{\rm sw}c^2}
\label{eq:Gt}
\end{eqnarray}
where $\Gamma$ is the bulk Lorentz factor, $L'_{\rm ra}$ is the comoving radioactive power, $L'_e$ is the comoving radiated bolometric luminosity, $E'_{\rm int}$ is the comoving internal energy, ${\cal D}=1/[\Gamma(1-\beta)]$ is the Doppler factor. The ejecta will interact with the interstellar medium to generate an external shock, with the shock swept mass from the interstellar medium (with density $n$)  estimated as $M_{\rm sw}=\frac{4\pi}{3}R^3nm_p$, where $R$ is the radius of the ejecta \citep{nakar11,gao13}.

The variation of $E'_{\rm int}$ could be expressed as \cite[e.g.][]{kasen10,yu13}
\begin{eqnarray}
{dE'_{\rm int}\over dt'}=L'_{\rm ra} -L'_{\rm e}
-\mathcal P'{dV'\over dt'}.
\label{eq:Ep}
\end{eqnarray}
where $\mathcal P'=E'_{\rm int}/3V'$ is the radiation dominated pressure. The comoving volume evolution can be fully addressed by
$dV'/dt'=4\pi R^2\beta c$ together with $dR/dt=\beta c/ (1-\beta)$ .

The comoving radioactive power and radiated bolometric luminosity could be expressed as
\begin{eqnarray}
L'_{\rm ra}=4\times10^{49}M_{\rm ej,-2}\left[{1\over2}-{1\over\pi}\arctan \left({t'-t'_0\over
t'_\sigma}\right)\right]^{1.3}~\rm erg~s^{-1},
\label{eq:Lrap}
\end{eqnarray}
where $t'_0 \sim 1.3$ s and $t'_\sigma \sim 0.11$ s \citep{korobkin12} and
\begin{eqnarray}
L'_e=\left\{
\begin{array}{l l}
  {E'_{\rm int}c\over \tau R/\Gamma}, & \tau>1, \\
  {E'_{\rm int}c\over R/\Gamma}, &\tau<1,\\ \end{array} \right.\
  \label{eq:Lep}
\end{eqnarray}
where $\tau=\kappa (M_{\rm ej}/V')(R/\Gamma)$ is the optical depth of the ejecta with $\kappa$ being the opacity \citep{kasen10,kotera13}. 

Assuming a blackbody spectrum for the thermal emission of the merger-nova, the observed flux for a given frequency $\nu$ could be calculated as
\begin{eqnarray}
F_{\nu}={1\over4\pi D_L^2}{8\pi^2  {\cal D}^2R^2\over
h^3c^2\nu}{(h\nu/{\cal D})^4\over \exp(h\nu/{\cal D}kT'_{\rm eff})-1},
\end{eqnarray}
where $D_L$ is the luminosity distance, $T'_{\rm eff}=(E'_{\rm int}/aV'\max(\tau,1))^{1/4}$ is the effective temperature, $h$ is the Planck constant, $k$ is the Boltzmann constant and $a$ is the radiation constant. Here we adopt the luminosity distance of NGC 4993 (e.g., $D_L=39.472~\rm Mpc$ \cite{kopparapu08}) as the luminosity distance of the optical/IR counterpart of GW170817.

With the above formulae, we can calculate the theoretical light curves for both the ``blue'' and ``red'' kilonova components. With the combination of these two components, we fit the multi-wavelength observational data of the optical/IR counterpart of GW170817, as collected by Villar et al. (2017). We use the ``emcee" python module based on Markov Chain Monte Carlo \citep{foreman-Mackey13} to constrain the ejecta parameters by performing a maximum likelihood $\chi^2$ fit. The data and best fit light curves are shown in the left panel of Figure 1. The right panel of Figure 1 is the contour plot for the parameter constraints. For the purpose of this work, we focus on ejecta mass, which are constrained as $M_{\rm ej,red}=0.045^{+0.018}_{-0.017}~\rm{M_{\odot}}$ and $M_{\rm ej,blue}=0.019^{+0.009}_{-0.005}~\rm{M_{\odot}}$ from our model fitting. Many groups have carried out data fitting invoking different degrees of sophistication in terms of modeling, e.g. by invoking the distribution of mass with velocity of the ejecta, detailed modeling the thermalization efficiency $\epsilon_{\rm th}$, or tracing the radioactive reactions so as to obtain the opacity evolution history, etc. \cite[][for a review]{metzger17b}. Our fitting results are fully consistent with others' results\footnote{In the literature, the optical/IR counterpart following GW170817 is interpreted as the combination of a ``blue'' and a ``red'' kilonova emission component. The early ``blue'' emission phase requires a lanthanide-free disk wind ejecta with mass $M_{\rm ej,blue}\approx0.01-0.04~\rm{M_{\odot}}$ and initial speed $\beta_{\rm i, blue}\approx0.2-0.3$ \citep{kasen17,cowperthwaite17,chornock17,kilpatrick17,villar17}. The late ``red'' emission phase requires a lanthanide-rich dynamical ejecta with mass $M_{\rm ej,red}\approx0.03-0.05~\rm{M_{\odot}}$ and initial speed $\beta_{\rm i, red}\approx0.1-0.2$ \citep{evans17,drout17,nicholl17,smartt17,arcavi17,kasen17,cowperthwaite17,kilpatrick17,villar17}.}. It is worth noticing that at this stage, we have no prior assumption that which component is from the dynamical ejecta.

\section{Mass Ratio Constraint}

\begin{figure*}[t]
\begin{center}
\begin{tabular}{ll}
\resizebox{80mm}{!}{\includegraphics[]{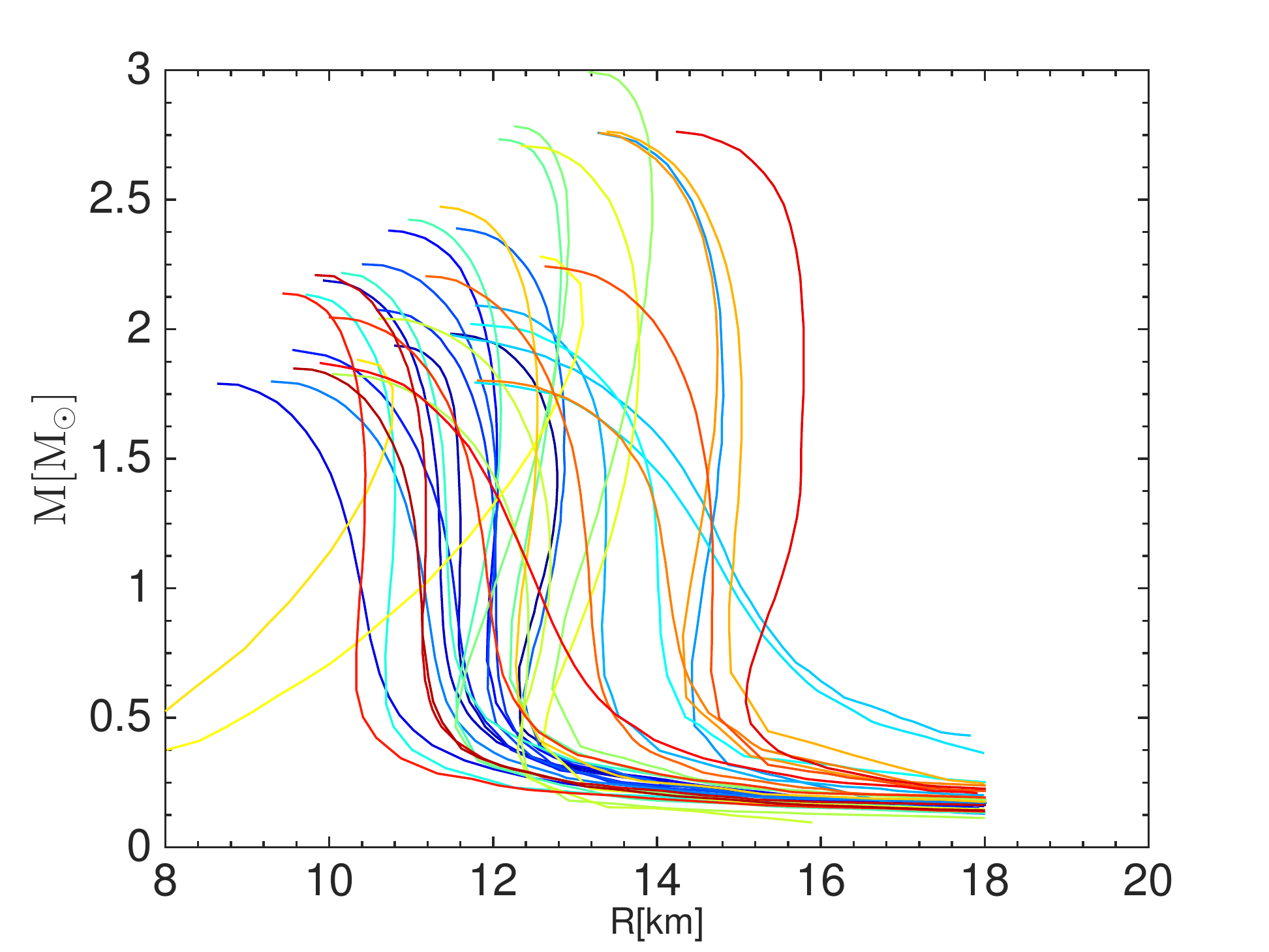}} &
\resizebox{80mm}{!}{\includegraphics[]{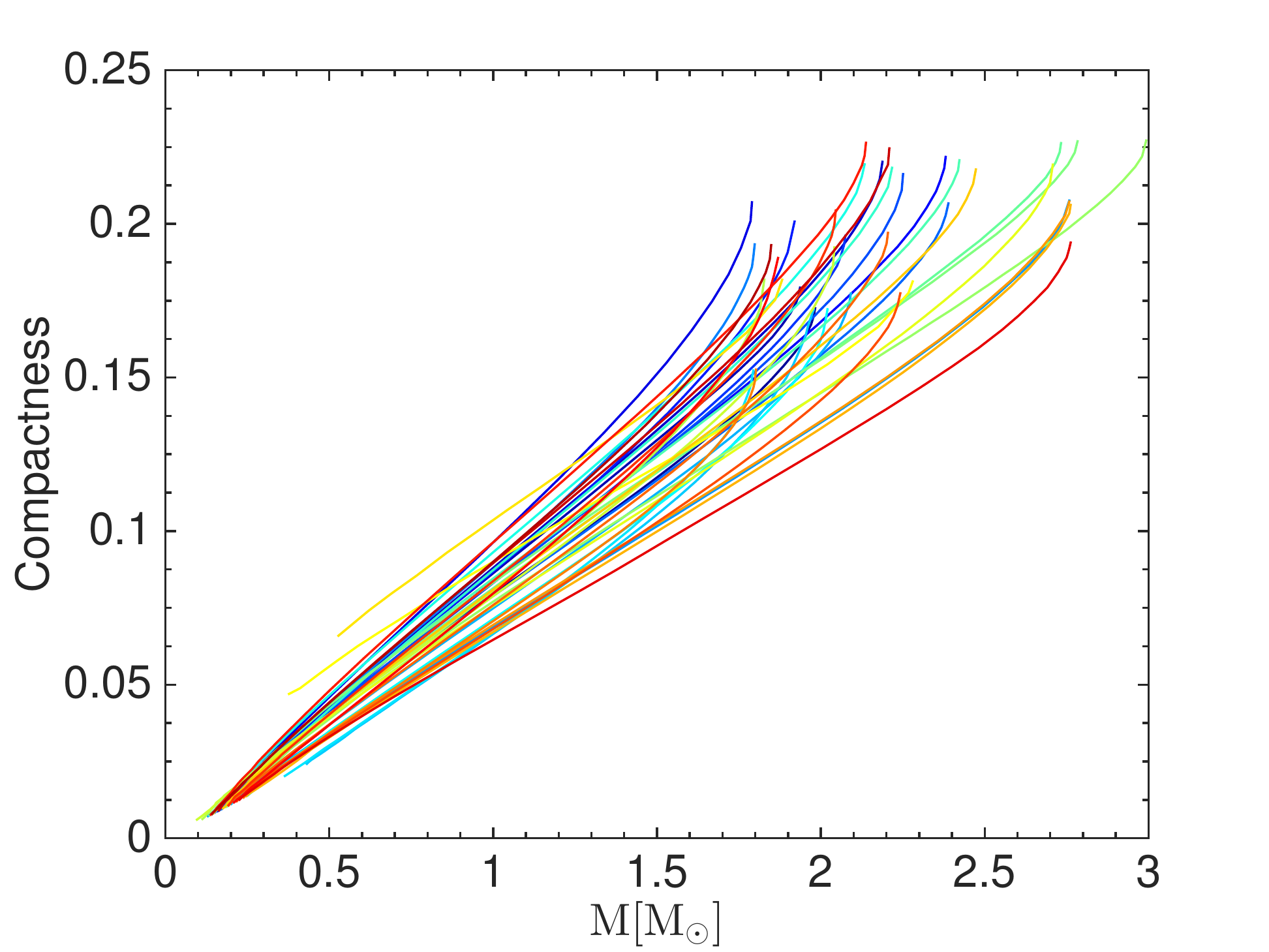}} \\
\end{tabular}
\caption{Mass-radius relationship (left panel) and mass-compactness relationship (right panel) for different neutron star equations of state.}
\label{fig:MRC}
\end{center}
\end{figure*}

A large number of groups have studied the merger process of NS-NS systems with numerical relativity simulations \cite[][and reference therein]{baiotti17}. It becomes a common approach to extract information from numerical relativity simulations about the properties of dynamical ejecta from the system. Most recently, \cite{dietrich17a} combined the published works from different groups and obtained a catalog of simulation results to derive fitting formulas for important dynamical ejecta quantities, such as the mass, kinetic energy, and velocity of the ejecta. For the purpose of this work, we employed their fitting formula for the dynamical ejecta mass:
\begin{eqnarray}
    \frac{M_{\rm ej}}{10^{-3}M_{\odot}}=&&\left[a\left(\frac{M_2}{M_1}\right)^{1/3}\left(\frac{1-2C_1}{C_1}\right)
    +b\left(\frac{M_2}{M_1}\right)^{n}\right]M_1^*\\ \nonumber
    &+&\left[a\left(\frac{M_1}{M_2}\right)^{1/3}\left(\frac{1-2C_2}{C_2}\right)
    +b\left(\frac{M_1}{M_2}\right)^{n}\right]M_2^*\\ \nonumber
    &+&c[(M_1^*+M_2^*)-(M_1+M_2)]+d
\end{eqnarray}
where $M_{1,2}$ and $M_{1,2}^*$ are the gravitational and baryonic mass of the two NSs, $C_{1,2}$ are their respective compactness, and the fitting coefficients are $a=-1.357$, $b=6.113$, $c=-49.434$, $d=16.114$ and $n=2.548$. For a given NS EoS, $C_{1,2}$ are determined by $M_{1,2}$. In figure \ref{fig:MRC}, we plot the mass-radius and mass-compactness relations for different NS EoSs employed in this work. Moreover, the baryonic mass $M_{1,2}^*$ are approximately related to the gravitational mass through the relation\footnote{Recently, \cite{coughlin17} proposed a new relation between the quotient of baryonic and gravitational mass $M^*/M$ and the compactness $C$ of a single neutron star, i.e., $M^*/M=1+aC^n$, where $a=0.8858$ and $n=1.2082$. We note that for the EoSs employed in this work, the two relations only deviate from each other by $(1-7) \%$.} $M^*=M+0.075M^2$, where the mass are in unit of the solar mass $M_\odot$ \citep{timmes96}. 

\begin{figure*}[t]
\begin{center}
\begin{tabular}{ll}
\resizebox{90mm}{!}{\includegraphics[angle=270]{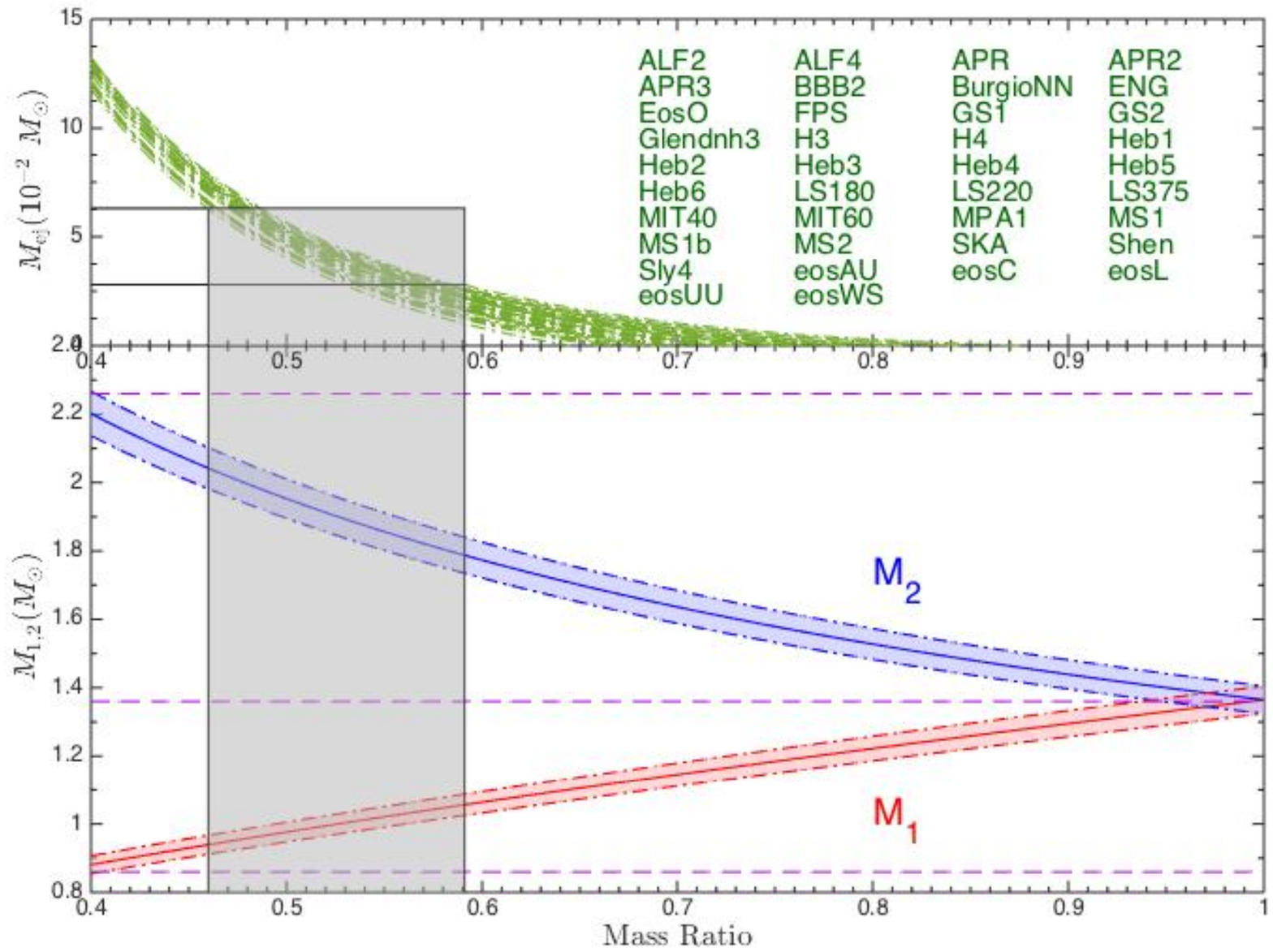}} &
\resizebox{90mm}{!}{\includegraphics[angle=270]{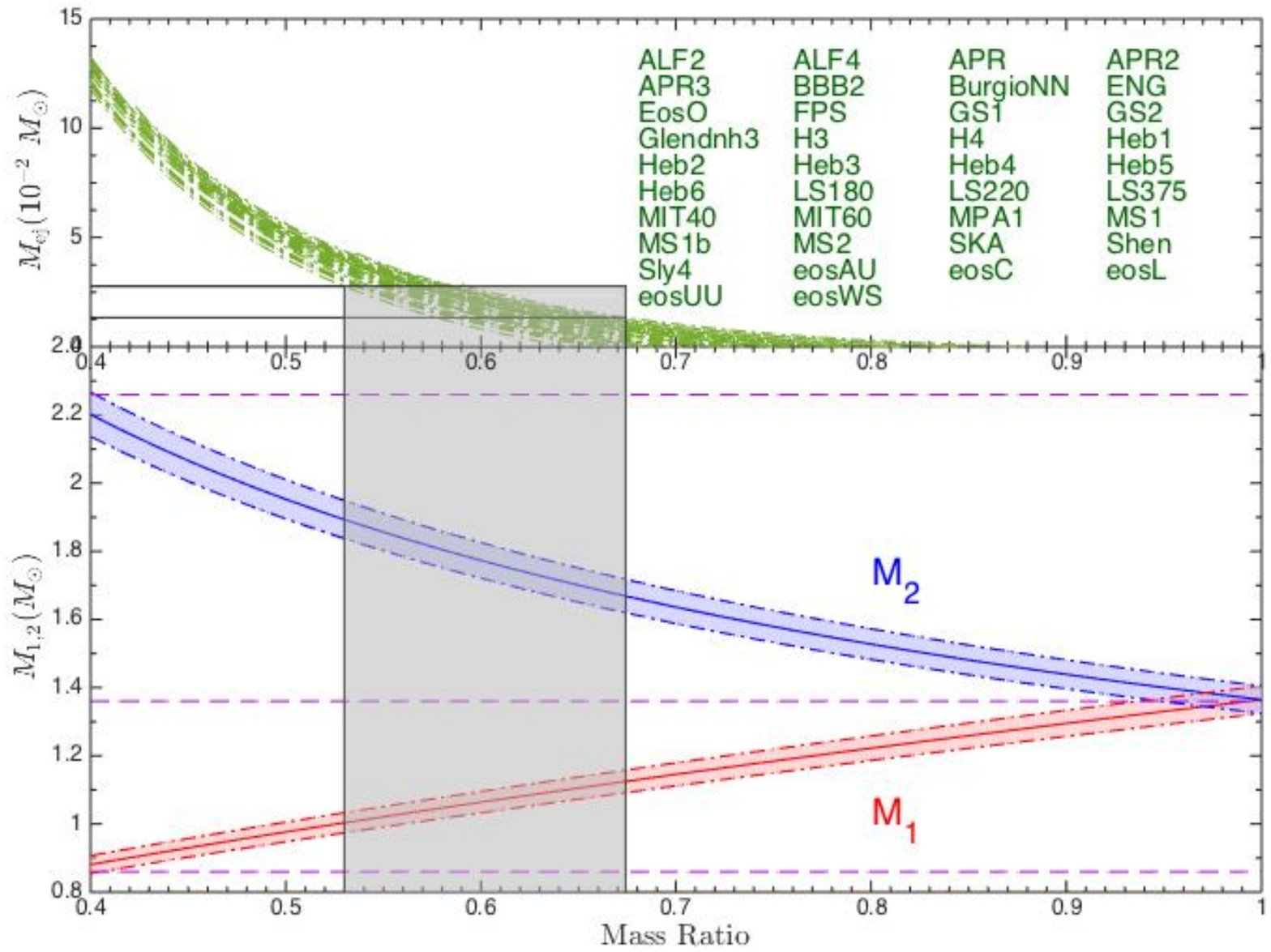}} \\
\end{tabular}
\caption{Constraint on the mass ratio of the two compact objects for GW170817. The left panel is for the case when the red ``kilonova'' being powered by the dynamical ejecta and the right panel is for the case when the blue ``kilonova'' being powered by the dynamical ejecta. For each panel, the top sub-figure shows how dynamical ejecta mass varying with the mass ratio for different EoS. Bottom sub-figure shows how mass of compact objects varying with the mass ratio for GW170817. The grey shadowed region presents the constraint results on the mass ratio.}
\label{fig:Ratio}
\end{center}
\end{figure*}

For an NS-NS system, given the chirp mass ($M_{c}$) and the mass ratio ${\cal R}=M_2/M_1$, the gravitational mass of the NSs could be derived as
\begin{eqnarray}
M_1=M_c\left(\frac{1+{\cal R}}{{\cal R}^3}\right)^{1/5},~~~M_2=M_1*{\cal R}.
\end{eqnarray}
For GW170817, the chirp mass of the system is precisely constrained to $1.188^{+0.004}_{-0.002}~\rm{M_{\odot}}$. On the other hand, for a wide range of NS EOSs, the dynamical ejecta mass is essentially determined by the mass ratio of the two NSs. In order to be as comprehensive as possible, we employ 38 representative microphysical EoSs\footnote{Notice that some EoSs employed here might be excluded by the known neutron star masses, such as the well-measured neutron star mass of $\sim2\rm{M_{\odot}}$ \citep{freire08}.}, which are derived in \cite{bauswein12} and \cite{bauswein13}, within different theoretical frameworks, including nonrelativistic and relativistic nuclear energy-density functionals, Brueckner-Hartree-Fock calculations, and phenomenological models, such as the liquid-drop model \cite[][for details]{bauswein12,bauswein13}. In Figure \ref{fig:Ratio}, we plot how the mass of the dynamical ejecta varies with the mass ratio for different EoSs. 

It is generally assumed \citep[e.g.][]{metzger14} that the red kilonova component following GW170817 is powered by the dynamical ejecta and the blue component is powered by the disk wind ejecta, so that $M_{\rm ej,red}$ should correspond to the dynamical ejecta mass $M_{\rm ej,dyn}$. In this case, we have $M_{\rm ej,dyn}=0.045^{+0.018}_{-0.017}~\rm{M_{\odot}}$. Considering all EOS, the mass ratio of GW170817 system is constrained in a tight range of $0.46-0.59$. Therefore, the mass of two NSs could be constrained to $1.73~{\rm M_{\odot}}<M_1<2.11~{\rm M_{\odot}}$ and $0.90~{\rm M_{\odot}}<M_2<1.09~{\rm M_{\odot}}$ respectively.

Alternatively, it is possible that the blue kilonova following GW170817 might be of a dynamical origin, while the red component might be from the disk wind \cite[e.g.][]{kasen17,nicholl17}. If this is the case, according to our fitting result, one would have $M_{\rm ej,dyn}=0.019^{+0.009}_{-0.005}~\rm{M_{\odot}}$. In this case, the mass ratio would be constrained in a tight range of $0.53-0.67$. Therefore, the mass of two NSs could be constrained to $1.61~{\rm M_{\odot}}<M_1<1.96~{\rm M_{\odot}}$ and $0.97~{\rm M_{\odot}}<M_2<1.16~{\rm M_{\odot}}$ respectively. 

Even the two different interpretations lead to somewhat different ranges of the mass ratio, these constrained ranges in any case are much narrower than that derived from the GW data alone. The component masses are therefore constrained to narrower ranges than using the GW data alone. Our results suggest that multi-messenger data are helpful to reach better constraints on the binary properties.

\section{Conclusion and Discussion}

The recent observations of GW170817 and its EM counterpart have opened the era of GW-led multi-messenger astronomy. It is of great interest to investigate how to make full use of multi-messenger information to better study the physical properties of the GW sources. Based on the results from numerical relativity simulations, a rough empirical relation between the dynamical ejecta mass and the mass ratio of the NS binary has been established. Before the detection of GW170817, it has been proposed that such a relation could be used to explore NS-NS binary parameters from kilonova detections \citep{coughlin17}. The LIGO-Virgo collaboration have employed such a relation to estimate the dynamical ejecta mass by using the GW measurements of GW170817 and used the resulting ejecta mass to estimate its contribution (without the inclusion of the disk wind ejecta) to the corresponding kilonova light curves from various models, but they did not directly invoke  electromagnetic observations \citep{abbott17e}. In this work, we fit the multi-wavelength data of the optical/IR transient following GW170817 with ``red''+``blue'' kilonova emission, and obtain the ejecta mass for both components ($M_{\rm ej,red}=0.045^{+0.018}_{-0.017}~{\rm M_{\odot}}$ and $M_{\rm ej,blue}=0.019^{+0.009}_{-0.005}~\rm{M_{\odot}}$). This henceforth leads to a more stringent constraint on the mass ratio of the NS binary. If the red component mass corresponds to the dynamical ejecta mass (Case I), the mass ratio would be constrained to the range of ($0.46-0.59$), which provides a much better estimation of the masses of the two merging NSs ($1.73~{\rm M_{\odot}}<M_1<2.11~{\rm M_{\odot}}$ and $0.90~{\rm M_{\odot}}<M_2<1.09~{\rm M_{\odot}}$). If, on the other hand, the blue component mass corresponds to the dynamical ejecta mass (Case II), the mass ratio would be constrained to the range of ($0.53-0.67$), which provides an estimation of the masses of the two merging NSs ($1.61~{\rm M_{\odot}}<M_1<1.96~{\rm M_{\odot}}$ and $0.97~{\rm M_{\odot}}<M_2<1.16~{\rm M_{\odot}}$). Overall, such a multi-messenger approach could make a constraint on the mass ratio of GW170817 system to the range of $0.46-0.67$, and a constraint on the individual NS mass as $1.61~{\rm M_{\odot}}<M_1<2.11~{\rm M_{\odot}}$ and $0.90~{\rm M_{\odot}}<M_2<1.16~{\rm M_{\odot}}$. If our interpretation is correct, some important implications could be inferred.

\begin{itemize}
\item In the era of GW-led multi-messenger astronomy, a comprehensive analysis of multi-messenger signals would shed light on both GW and EM studies. As we have shown, if GW170817 indeed originates from a double NS merger, the follow up optical/IR observations could help to place a much more stringent constraint on the mass ratio of the two NSs than the constraint from the analysis using pure GW signals. The results are barely affected by the unknown EoS of NSs. 
\item The lower mass NS in GW170817 event is close to or even smaller than $1.0~\rm M_{\odot}$, falling into (or even below) the lower portion of the mass distribution of known NSs \citep{lattimer12}. If this is true, this result would raise an important question for stellar evolution models regarding how to generate such a low mass neutron star. 
\item With the results from the GW signal analysis, the mass ratio is bound to $0.7-1$ under the low dimensionless NS spin prior ($\chi<0.05$), and is bound to $0.4-1$ under the high spin prior ($\chi<0.89$). 
The mass ratio value from our result is smaller than 0.67, which is contradictory to the low spin prior, inferring that the two NSs involving in the GW170817 event tends to have relatively high dimensionless spin, which is unexpected based on the observed galactic NS binary population \citep{abbott17c}. More diverse properties for extra-galactic binary NS systems might exist and is required for further investigations. 
\end{itemize}

It is worth noticing that the empirical relation between the dynamical ejecta mass and the mass ratio of two NSs has an average of uncertainty of $\sim 72\%$ \citep{dietrich17a}, which is the same order as the numerical uncertainty of the numerical relativity simulations. For the sake of conservatism, we artificially extend our error bar for $M_{\rm ej,dyn}$ by a factor of 1.72. For Case I, the mass ratio could be constrained into the range of $0.44-0.66$, so that the mass of two compact objects could be constrained to $1.64~{\rm M_{\odot}}<M_1<2.15~{\rm M_{\odot}}$ and $0.89~{\rm M_{\odot}}<M_2<1.14~{\rm M_{\odot}}$ respectively. For Case II, the mass ratio could be constrained into the range of $0.51-0.71$, so that the mass of two compact objects could be constrained to $1.58~{\rm M_{\odot}}<M_1<1.98~{\rm M_{\odot}}$ and $0.96~{\rm M_{\odot}}<M_2<1.19~{\rm M_{\odot}}$ respectively. The uncertainty of the relation does not affect the main results. More numerical relativity simulations are required in the future for reducing the uncertainty. 

The empirical relations proposed in \cite{dietrich17a} did not involve the effects of the spin of individual NSs, because the numerical relativity fits in \cite{dietrich17a} are based on a sample of representative simulations, where the effects of the spin of individual NSs are essentially not considered. Recent simulations suggested that the spin of the individual NSs may either increase or decrease the ejecta mass, depending on the adoption of EoS and the initial configuration or other properties of the NS-NS system \citep{kastaun17,dietrich17b}. However, limited EoSs and initial configurations have been investigated in these references. At this stage, it is difficult to quantitatively justify the uncertainty caused by this effect to the fitting results in \cite{dietrich17a}, so as to our results. Specific numerical simulations are required in the future to investigate this issue.

It is possible that both the blue and red components have contributions from both the dynamical ejecta and disk wind components. However, each component should be dominated by one type of ejecta. Otherwise (e.g. equal contributions from both components or one component dominating in both the blue and red components), it would be more challenging to interpret the apparent diverse properties of the two emission components.

\acknowledgments

We thank Brian Metzger for helpful discussion, and an anonymous referee and Zigao Dai for important comments. This work is supported by the National Basic Research Program (973 Program) of China (Grant No. 2014CB845800), the National Natural Science Foundation of China under Grant No. 11722324, 11603003, 11633001 and 11690024, and the Strategic Priority Research Program of the Chinese Academy of Sciences, Grant No. XDB23040100.

{}

\end{document}